\documentclass[10pt,prl,twocolumn,superscriptaddress,preprintnumbers,amsmath,amssymb,floatfix]{revtex4-2}

\usepackage{amsmath}
\usepackage{mathrsfs}

\DeclareFontFamily{U}{calligra}{}
\DeclareFontShape{U}{calligra}{m}{n}{<->callig15}{}

\newcommand{\calE}{{\!\!\text{\usefont{U}{calligra}{m}{n}E}\,\,}}
\newcommand{\calB}{{\!\!\text{\usefont{U}{calligra}{m}{n}B}\,\,}}

\usepackage[T1]{fontenc}
\usepackage{currvita}

\usepackage{graphics,amssymb,amsmath,epsfig,color}
\usepackage{graphicx}
\usepackage{bm}

\usepackage{dcolumn}% Align table columns on decimal point
\usepackage{bm}% bold math
\usepackage{float}
\usepackage{natbib}

\usepackage[export]{adjustbox}

\usepackage{braket}
\setcitestyle{square}

\usepackage{graphicx}% Include figure files
\usepackage{dcolumn}% Align table columns on decimal point
\usepackage{bm}% bold math
\usepackage{float}
\usepackage{natbib}
\setcitestyle{square}

\usepackage{xcolor}

\usepackage{hyperref}
\hypersetup{colorlinks=true, linkcolor=blue, citecolor=blue, urlcolor=blue}

\begin{document}

\title{A phonon-driven mechanism for an emergent and reversible chirality in crystals}

\author{Mauro Fava}
 \affiliation{Physique Th\'eorique des Mat\'eriaux, QMAT, Universit\'e de Li\`ege, B-4000 Sart-Tilman, Belgium}
 
\author{Emma McCabe}
\affiliation{Department of Physics, Durham University, South Road, Durham, DH1 3LE, U. K.}

\author{Aldo H. Romero}
\affiliation{Department of Physics and Astronomy, West Virginia University, Morgantown, WV 26505-6315, USA}

\author{Eric Bousquet}
\affiliation{Physique Th\'eorique des Mat\'eriaux, QMAT, Universit\'e de Li\`ege, B-4000 Sart-Tilman, Belgium}

\date{\today}

\begin{abstract}
We demonstrate through first-principles calculations applied to the K$_{3}$NiO$_{2}$ crystal that a structural phase transition from an achiral to a chiral phase can be mediated by a degenerate soft phonon mode and controlled by pressure and epitaxial strain. 
Breaking such degeneracy with an electric field generates a competition between enantiomorphic, polar, and orthorhombic displacements. 
Originated by the interaction between spontaneous chiral and induced polar and axial modes, an optimal parameter window for converting the handedness of the system into its opposite kind 
is observed. 
\end{abstract}

\maketitle

Periodic solids are chiral if their structures are described by space groups containing only proper symmetry operations that map, for example, a right-handed coordinate system onto a right-handed one, and not onto a left-handed one (``operations of the first kind''),~\cite{Nespolo2018} otherwise they are achiral~\cite{flack}.
In recent years there has been a growing interest in investigating certain chiral effects (both in real- and spin-space) in crystals such as the Chiral Induced Spin Selectivity (CISS) and spin spirals~\cite{Yang2021,Evers2022}, polar ~\cite{RevModPhys.95.025001} and magnetic skyrmions~\cite{PhysRevLett.87.037203,Fert2017}, topological states~\cite{Chang2018,schroter2020} and chiral phonons~\cite{PhysRevLett.115.115502,Ishito2023,Ueda2023} to mention a few.
Noticeably, the crystallization process fixes the chirality and associated properties in most of these instances.
While the concept of spontaneous symmetry breaking has so far been a fertile ground for important discoveries in condensed matter physics~\cite{beekman2019introduction}, with phenomena such as ferroelectricity, magnetism, or superconductivity, to date, it has not been applied to study the possibility of chirality spontaneously emerging from an achiral solid phase. 
Nevertheless, a chiral irreducible representation (IRREP) or order parameter can be formally defined~\cite{hlinka2014}, so that a group-subgroup relationship of the type achiral $\rightarrow$ chiral between two phases is a physical possibility~\cite{chiroaxial_Erb_Hlinka}. 
Despite the recent experimental reports about the spontaneous formation of gyrotropic domains observed in certain materials~\cite{hayashida2021b,Kimura2016,PhysRevB.102.235127,Hayashida2021,Hayashida2022, PhysRevB.109.024113,fava2023ferroelectricity} below a critical temperature, 
many questions remain to be answered about their physical origin, properties and the way the can be controlled. Importantly, it is not clear if the handedness of a chiral material can be effectively converted into the opposite kind via the sole application of external fields.

In this work, we attempt to address these issues 
by exploring the achiral to chiral enantiomorphic phase transition observed in K$_3$NiO$_2$ (KNO)~\cite{djurivs2012k3nio2}. 
Using symmetry analysis and  density functional theory (DFT) calculations (see the Supplementary file~\cite{supp} for technical details and  references~\cite{Kresse1999,Perdew1996,PhysRevB.57.1505,PhysRevB.71.035105,isodistort,amplimode,gonze_2020,pseudodojo,NAO_expt,phonopy1,phonopy2} there included), we show that a soft zone boundary phonon mode triggers the emergence of chirality in this material with four well-like shapes in the energy landscape. 
We then explore how this structural chiral energy surface is enhanced or reduced by pressure or strain, and going further, we analyze how the chiral distortion can couple non-linearly with infrared (IR) and axial Raman active modes activated by an electric field.
We prove that direct control over these polar and axial distortions is crucial to attain a permanent enantiomeric excess in a hysteresis-like manner and after said external field is removed.
We discuss how this mechanism can be generally applied to several families of materials, with having one degenerate instability generating both opposite enantiomers as the sole requirement.

\begin{figure}
     \includegraphics[width=0.45\textwidth]{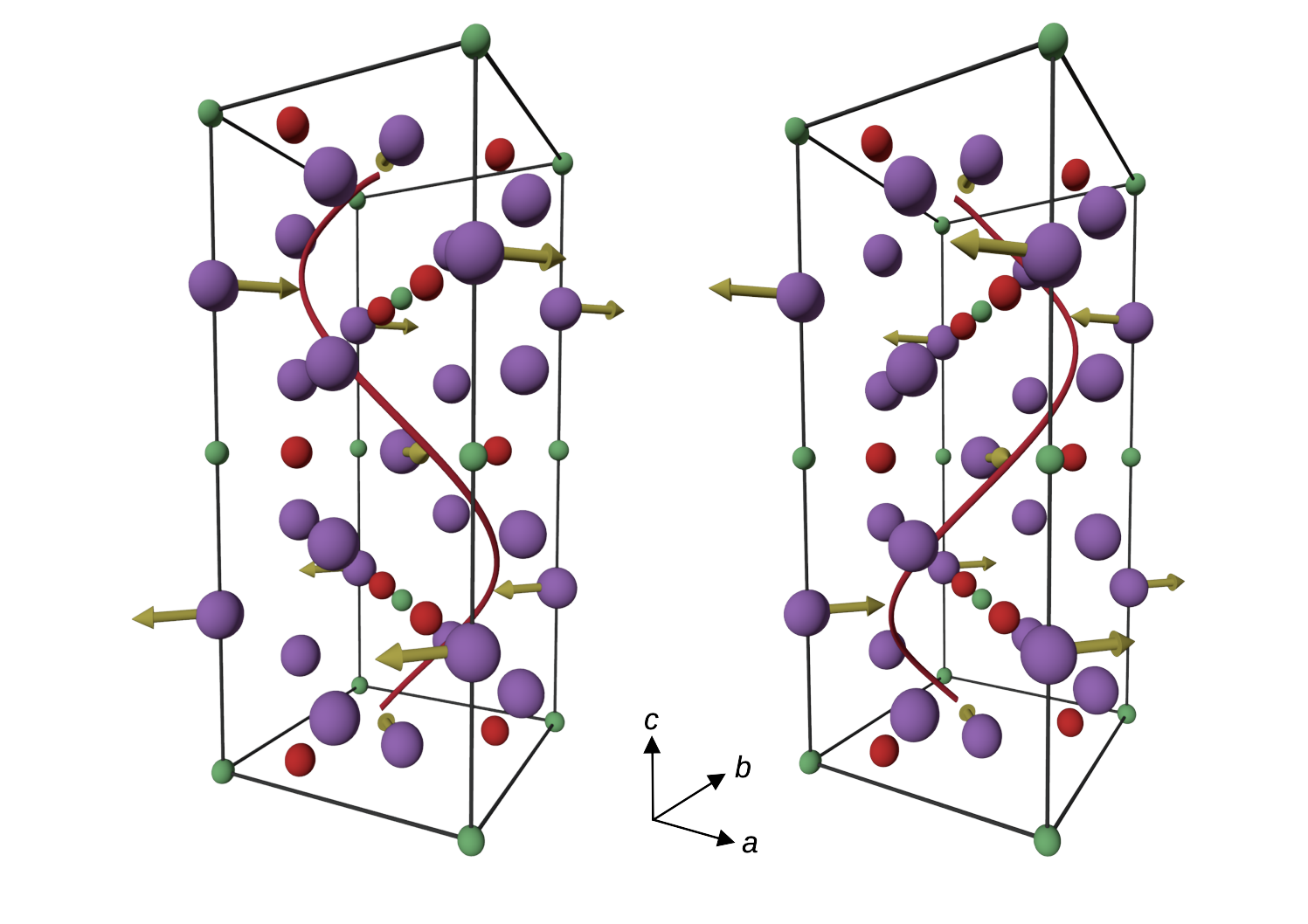}
     \caption{Scheme of the soft chiral phonon distortion in K$_{3}$NiO$_{2}$. The purple, green, and red spheres correspond to K, Ni, and O atoms. The ions are in the $P4_{2}/mnm$ configuration (1x1x2 supercell), while the golden arrows and the red spirals represent the $Z_{4}$ chiral ($P4_{1}2_12$ left , $P4_32_12$ right) displacement (for clarity, only the main K-motion is shown).}
     \label{fig:figure_1}
     \end{figure}

\begin{figure}
     \includegraphics[width=0.45\textwidth]{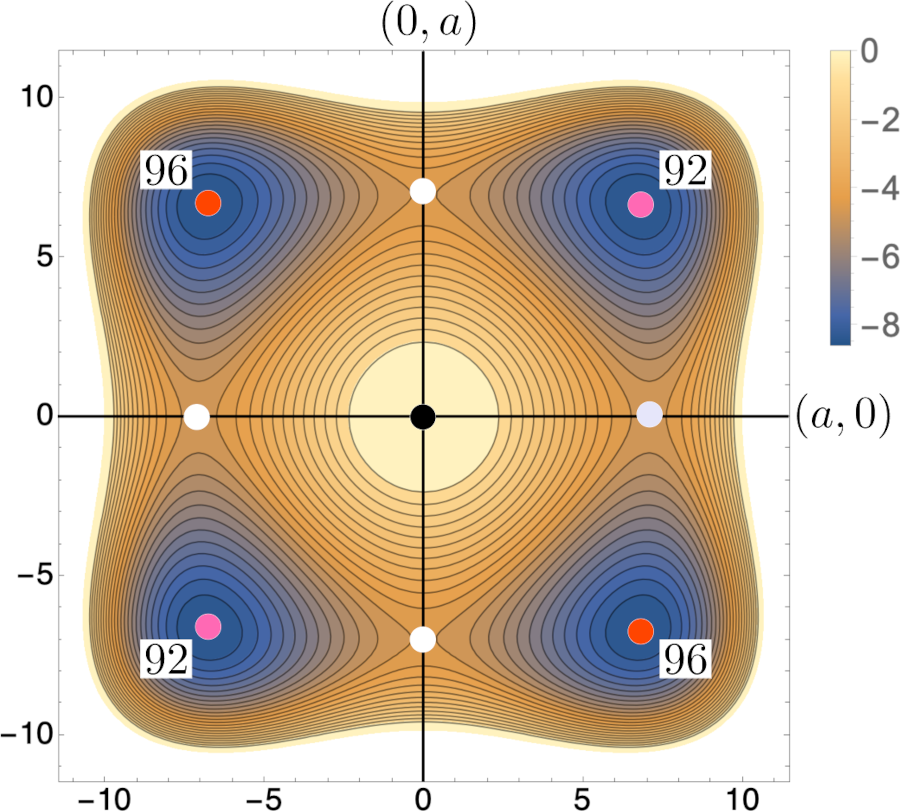}
     \caption{Schematic 2D projection of the surface energy given by condensing the two-dimensional $Z_4$ mode (amplitudes in arbitrary units). 
     The central black circle corresponds to the high symmetry $P4_2/mnm$ phase.      
     The  $Cmcm$ (no. 63) energy minimum points 
     are highlighted with white circles.
     Red and pink-filled circles indicate the degenerate ground state points in the $P4_32_12$ (no. 96) and $P4_12_12$ (no. 92) enantiomorphic phases, respectively.}
     \label{fig:energy_landscape}
     \end{figure}

The high symmetry phase of KNO adopts an achiral tetragonal structure described by space group symmetry $P4_{2}/mnm$ (no. 136). 
On cooling below $\sim$ 423 K, it undergoes a first-order phase transition to a low-temperature enantiomorphic phase of either $P4_{1}$2$_{1}2$ (no. 92) or $P4_{3}$2$_{1}2$ (no. 96) symmetry~\cite{djurivs2012k3nio2}. Our analysis of the experimentally reported structures~\cite{djurivs2012k3nio2} hence shows that the corresponding symmetry-lowering structural deformation decomposes into a main zone boundary (Z = 0,0,1/2) and a secondary pseudoscalar zone centre symmetry adapted modes (SAMs) with Z$_4$ and A$_{1u}$ representation, respectively. 
We perform our calculations with the collinear spins and the Hubbard-U correction (U = 4.2 eV, computed via the Cococcioni's method~\cite{PhysRevB.71.035105}) on Ni ions (required to open a band gap)~\cite{supp} and compute~\cite{phonopy1,phonopy2} the phonon spectrum in the $P4_2/mnm$ phase. We identify, consistently with the previous SAM analysis, a doubly degenerate soft mode at the Z point with frequency $\omega_\text{Z}$ = 1.35i THz and Z$_4$ irreducible representation.
While other soft modes (at $\Gamma$, X and M points~\cite{supp}) are detected, the Z mode leads to the most stable low-symmetric configuration, as confirmed by the experiment~\cite{DURIS2012,djurivs2012k3nio2,djurivs2012syntheses}. The computed $Z_{4}$ distortion alone induces the chiral phases and chiefly involves the in-plane displacement (0.4 \AA/ion) of potassium ions from their $P4_2/mnm$ position at the Ni-O-K-O linear chains to form a handed 4$_{\{1,3\}}$ spiral as in fig.~\ref{fig:figure_1}. As a consequence, the chain O - K - O bond angle becomes smaller than 180$^\circ$ whilst the NiO$_{2}^{3-}$ units remain close to linear, in agreement with experimental structure reports~\cite{djurivs2012k3nio2, DURIS2012, Moeller1995}.
A secondary A$_{1u}$ SAM, observed upon optimization of the chiral structures, is instead solely constituted by a small displacement (0.006 \AA/ion) of tetrahedrally coordinated K atoms along the $c$ axis (see supplementary section III~\cite{supp}). 
Since the Z$_4$ soft boundary mode is doubly 
degenerate, we can build arbitrary linear combinations of eigenstates and calculate the energy as a function of their amplitude as illustrated in fig.~\ref{fig:energy_landscape}. 
The free energy F - which includes the chiral $\phi_Z = (\phi_a,\phi_b)$  SAMs amplitudes - thus reads $F(\mathbf{\phi}_{Z}) = \alpha|\mathbf\phi_{Z}|^2 + \beta_{1}|\mathbf\phi_{Z}|^4 + \beta_{2}(\phi_a^4 + \phi_b^4)$.
In particular, while the diagonal directions ($\phi_a$,$\phi_a$) and ($\phi_a$, -$\phi_a$)  are associated with the realization of the $P4_{1}$2$_{1}$2 and $P4_{3}$2$_{1}$2 enantiomorphic domains, respectively, different settings lead to structures with different symmetries. More specifically, the ($\phi_a$,0) and (0,$\phi_b$) directions produce degenerate orthorhombic phases with $Cmcm$ (no. 63) symmetry, which are achiral and optically inactive, while arbitrary ($\phi_a$,$\phi_b$) (with $\phi_a\neq$ 0, $\phi_b\neq$ 0 and $\phi_a\neq\pm\phi_b$) combinations generate systems with $C222_{1}$ symmetry.

\begin{figure*}[htb!]
\centering
     \includegraphics[width=1.0\textwidth]{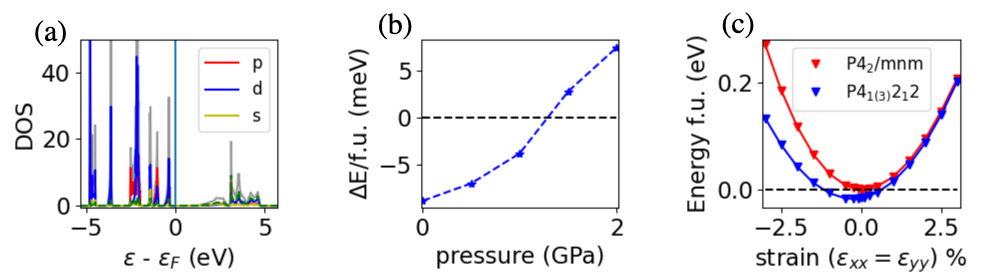}
     \caption{(a) Partial electronic density of states near the Fermi level (P4$_{2}$/mnm phase). The dashed green curve represents K states and the grey curve the total DOS. (b) and (c) $\Delta$E($P4_2/mnm\rightarrow P4_{1(3)}2_12$) behavior as a function of the hydrostatic pressure and epitaxial in-plane strain respectively.}
     \label{fig:DFT_quantities_merge}
     \end{figure*}

The electronic density of states (DOS) near the Fermi level (see Fig.~\ref{fig:DFT_quantities_merge}(a)) is mainly populated by weakly hybridized O-$2p$/Ni-$3d$ valence and K-$3d$ conduction states with the ground state in a low-spin configuration.
We find the density of states to be negligibly affected by the phase transition~\cite{supp}, which also agrees with the rather nominal calculated Born effective charges~\cite{supp} and possibly suggests a short-range interaction mechanism for the emergence of chirality, in agreement with its zone boundary nature~\cite{PhysRevLett.35.1767}. Concerning the behavior under mechanical perturbations, our results in Fig.~\ref{fig:DFT_quantities_merge}(b)) show that even a tiny amount of hydrostatic pressure (above a value between 1.0 and 1.5 GPa) hardens the chiral mode and destroys the phase transition. 
This hardening under pressure may explain the observed decrease of the transition temperature upon replacing K${^+}$ with larger Rb${^+}$ or Cs${^+}$ ions, namely in Rb$_{3}$NiO$_{2}$ (T$_{c}$ = 390 K) and Cs$_{3}$NiO$_{2}$ (achiral at room temperature)~\cite{DURIS2012}. 
Conversely, an even more nuanced effect can be observed in the presence of strain.
Indeed, the Z$_4$ mode is respectively hardened or softened if the applied in-plane $\epsilon_{xx}=\epsilon_{yy}$ lattice deformation is tensile or compressive, as shown in Fig.~\ref{fig:DFT_quantities_merge}(c). 
These results stem from a tension-compression asymmetry of the energy response
in both the high and low symmetry phases.
Nevertheless E($P4_2/mnm$) > E($P4_{1(3)}2_12$) always in the considered strain range (E is the energy, formula unit).
Noticeably we find the $\Gamma$ instability, which is polar ($E_u$ representation) and comes from short-range forces (due to the non-anomalous Born charges~\cite{supp}), to persist under tensile strain, with $\Delta$E($P4_2/mnm\rightarrow E_u$) = -2.3 meV upon $\epsilon_{xx}=\epsilon_{yy}$ = 3 \%.
While this value lies higher than the corresponding chiral state energy, it is possible that ferroelectricity may coexist with chirality at this or at a larger magnitude of $\epsilon_{xx}=\epsilon_{yy}$. 
We leave this possibility open for future exploration.

%%%%%%%%%%%%%%%%%%%%%%%%%%%%%%%%%%%%%%%%%%%%
\textit{Handedness conversion}. 
Synthesized samples prepared by an azide-nitrate route resulted in racemic twinned crystals with approximately equal amounts of each enantiomorph (Flack parameter = 0.53)~\cite{djurivs2012k3nio2}, illustrating the difficulty of preparing enantiomerically pure samples. 
At the same time, the Z$_4$ and A$_{1u}$ representations are coupled by the $\phi_a\phi_b\phi_{A_{1u}}$ invariant.
Therefore, any tensor $\mathbf{T}$ containing the pseudoscalar IRREP interacts with the chiral Z$_4$ mode through an invariant of the form $\phi_a\phi_b\mathbf{T}$, possibly allowing control over the handedness of the system. 
One such tensor may be obtained by taking the product between vector ($\mathbf{V}\equiv A_{2u} \oplus E_u$) and axial ($\mathbf{A} \equiv A_{2g} \oplus E_{g}$) quantities~\cite{PhysRevLett.129.116401}. 
Indeed, the representation of this product can be decomposed into its irreducible components as $\text{IRREP}(\mathbf{V}\otimes\mathbf{A}) = (A_{2u} \oplus E_{u})\otimes(A_{2g}\oplus E_{g}) = 2A_{1u} \oplus 2E_{u} \oplus A_{2u} \oplus B_{1u} \oplus B_{2u}$ (note that both the $A_{2u}\otimes A_{2g}$ and $E_u \otimes E_g$ products contain $A_{1u}$), thus revealing a pseudoscalar component. 
We will now construct a model - based on the interaction between chiral, polar, and axial modes - that breaks the $P4_12_12$/$P4_32_12$ degeneracy, realizes field-induced enantioselectivity and potentially creates a permanent enantiomeric excess.
Applying an electric field results in the activation of polar in-plane $(\phi^{a}_{IR},\phi^{b}_{IR})$ and out-of-plane $\phi_{A_{2u}}$ modes (with $E_u$ and $A_{2u}$ representations respectively). 
The axial Raman active modes $(\phi^{a}_{R},\phi^{b}_{R})$ with $E_g$ representation can be switched on as well via the trilinear $\sim\phi_{A_{2u}}(\phi^{a}_{R}\phi^{b}_{IR} - \phi^{b}_{R}\phi^{a}_{IR})$ interaction, following a scheme analogous to that of orthorhombic ErFeO$_{3}$~\cite{PhysRevLett.118.054101}.
In the following we will explore how these field-induced distortions couple with the Z$_{4}$ instability.
Along with the always present biquadratic couplings, the following terms are allowed: 
\begin{equation}\label{eq_flip}
V_{\text{flip}}\propto\phi_a\phi_b(\phi^{a}_R\phi^{a}_{IR} + \phi^{b}_R\phi^{b}_{IR})
\end{equation}
and
\begin{equation}\label{eq_cmcm}
    V_{\text{ortho}}\propto(\phi_a^2-\phi_b^2)\phi^{a}_{IR}\phi^{b}_{IR}.
\end{equation}

\begin{figure*}     
\centering
\includegraphics[width=1.0\textwidth]
{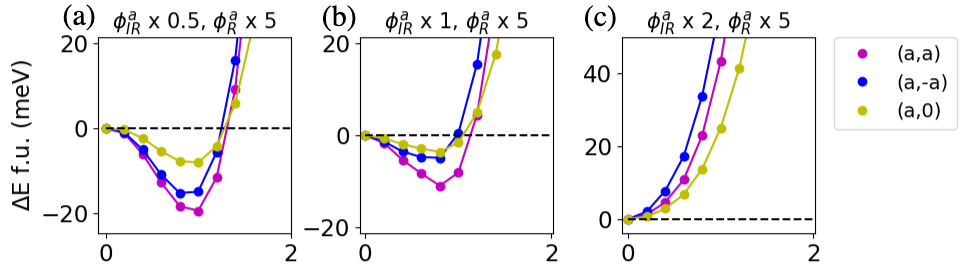}
     \caption{DFT-evaluated energy landscape as a function of the $\phi_Z$ amplitude in Na$_{3}$AuO$_{2}$ and in the presence of additional in-plane polar $E_u$ and axial $E_g$ independent distortions (continuous segments are a guide to the eye). 
     In this setting, $V_\text{ortho}$ = 0.
     The $E_g$ magnitude is fixed to be five times the  DFT-obtained value (via $A_{2u}$-$E_u$-$E_g$ coupling, see main text), 
     while the $E_u$ amplitude is scaled (0.5, 1 and 2 in figs. (b), (c) and (d) respectively). Fig.(b) shows the onset of a handedness flipping via crossing the $ P4_32_12\rightarrow Cmcm \rightarrow P4_12_12$ energy barrier.}
    \label{fig:NAO_field_no_strain}
     \end{figure*}

While $V_\text{flip}$ is an enantioselective mechanism that requires the activation of the $E_g$ modes, $V_\text{ortho}$ breaks the 
($\phi_a$,0)/(0,$\phi_b$) degeneracy of the $Cmcm$ achiral distortion. 
It is interesting to check if the energy barrier that separates the right- and left-handed domains can in practice be overcome by $V_\text{flip}$. 
This would lead to a permanent enantiomeric excess (i.e. surviving the $V_\text{flip}\rightarrow$ 0 condition).
To numerically test these ideas, we have extracted the relevant $E_g$ and $E_u$ modes via DFT structural relaxations in the presence of an electric field~\cite{PhysRevLett.89.117602}. 
Importantly and since including collinear magnetism and the Hubbard-U interaction proves too numerically cumbersome in the presence of an electric field, we replace the K and Ni positions with Na and Au respectively to form a filled $d$-shell paramagnetic Na$_{3}$AuO$_{2}$ (NAO) system with the same asymmetric unit of KNO and with a 2.1 eV DFT-computed band gap (without Hubbard correction). 
We note that although Na$_{3}$AuO$_{2}$ has been shown experimentally to adopt a slightly different structure
~\cite{Wagner1987}, our "toy model" (see the SI~\cite{supp} for the details of its construction) provides a helpful approximation of KNO. 
We apply an electric field $\calE$ = 4.63x$10^{4}$ kV/m along the [1,1,1] direction of the high symmetry phase of Na$_{3}$AuO$_{2}$, perform a structural relaxation and extract the mixed polar-Raman distortion $\equiv\ket{\delta_{PR}}$. 
Applied to the $P4_2/mnm\rightarrow P4_{1(3)}2_12$ energy landscape, we observe that the overall effect of $\ket{\delta_{PR}}$ is to stabilize the $Cmcm$ $(0,b)$ distortion while hardening all other orthogonal $Z_4$ directions ((a,$\pm$a) and (a,0)) as a consequence of the effect of $V_\text{ortho}$~\cite{supp}).
Importantly, no energy difference between 
mirror-symmetric structures is detected. 
These features are also robust in the presence of an additional $4/mmm\rightarrow mmm$ inducing compressive strain along the $x$-direction ($\epsilon_{xx}$ = 1 \%~\cite{supp}). 
The lack of right/left splitting can be explained if one realises that $V_\text{flip}\propto\epsilon_{xx}$ when $(\phi^{a}_{R},\phi^{b}_{R})\propto\phi_{A_{2u}}(-\phi^{b}_{IR},\phi^{a}_{IR})$~\cite{supp}. 
Thus while the gyrotropic properties of KNO could be affected by the field, the absence of any $P4_{1(3)}2_12\rightarrow P4_2/mnm \rightarrow P4_{3(1)}2_12$ or $P4_{1(3)}2_12\rightarrow Cmcm \rightarrow P4_{3(1)}2_12$ crossing prevents the realization of a permanent enantiomeric excess. Instead of exploiting the aforementioned $\sim\phi_{A_{2u}}(\phi^{a}_{R}\phi^{b}_{IR} - \phi^{b}_{R}\phi^{a}_{IR})$ nonlinear coupling in the presence of a constant bias along the [1,1,1] direction, we could use an axial external field - such as $\mathbf{\nabla}\times\calE$ or $\mathbf{\nabla}\times\mathbf{u}$~\cite{PhysRevLett.104.163901,PhysRevLett.129.116401} (where $\mathbf{u}$ defines an ionic displacement) - to switch the $E_g$ modes on. 
Assuming we can activate $E_u$ and $E_g$ phonons along the $(\phi^{a}_{IR},0)$ and $(\phi^{a}_{R},0)$ directions respectively ($\calE_z$ along the $c$ direction is set to zero), then $V_\text{ortho}\rightarrow$0 and the energy landscape may admit a sizable $V_\text{flip}$ if the axial mode is large enough. 
For sake of simplicity, we can assume to retain the polar-activated $E_g$ mode, although admitting a linear scaling of its amplitude to mimic an axial conjugate field. 
The effect of this strategy is shown in Fig.~\ref{fig:NAO_field_no_strain}(a,b,c). 
Fixing the $E_g$ amplitude, a small polar mode 
as in Fig.~\ref{fig:NAO_field_no_strain}(a)
($\phi^{a}_{IR}$ x 0.5) breaks the degeneracy between mirror equivalent enantiomers but cannot result in the crossing of the $P4_{3}2_12\rightarrow Cmcm \rightarrow P4_{1}2_12$ barrier.
On the other hand, increasing the polar scaling amplitude to 1 shows in Fig.~\ref{fig:NAO_field_no_strain}(b) that there exists an optimal window for the realization of a permanent enantiomeric excess, as measured by the difference between $P4_32_12$ and $P4_12_12$ concerning the $Cmcm$ distortion and $P4_2/mnm$ reference. 
Finally, a larger $E_u$ amplitude (e.g. $\phi^{a}_{IR}$ x 2) has the effect of stabilizing the polar $E_u$ state as reported in Fig.~\ref{fig:NAO_field_no_strain}(c).
Once the corresponding barrier has been crossed the system is expected to be remain scalemic (e.g. non racemic) in zero-field conditions. 
The overall handedness of the system can therefore be inverted, by changing the sign of the electric field direction and with a magnitude consistent with fig.~\ref{fig:NAO_field_no_strain}(b), consistently with a mechanism that is analogous to a hysteresis process.

\textit{Discussion}. The proposed model can be applied to several other classes of materials as long as chiral orders with opposite handedness emerge from one multidimensional IRREP, a condition present in many other known systems. For instance, a $\Delta_6$ (0,0,1/3) SAM induces a $P6_3/mmc\rightarrow P6_122, P6_522$ transition in hexagonal $ABX_3$ halide perovskites such as CsCuCl$_{3}$~\cite{Yamamoto2021,hirotsu1977}. At the same time, $AB_2X_4$ spinels such as MgTi$_2$O$_4$~\cite{Isobe2002,Schmidt2004,PRL_orbital_order,PRL_Peierls} along with SiO$_2$ cristobalite~\cite{PhysRevB.46.1,V_Dmitriev_1997,Leadbetter,Hatch1991} show a $Fd\Bar{3}m\rightarrow P4_32_12, P4_12_12$ displacive transition triggered by a soft X$_4$ (0,1,0) mode. 
These $\Delta_6$ and X$_4$ IRREPS span both enantiomorphic and orthorhombic displacements and calling
($\psi_a$,$\psi_b$) the order parameter ($\psi$ = $\Delta_6$, X$_4$), we can generally define V$_\text{flip} \equiv (\psi_{a}^2 - \psi_{b}^2)f_1(\mathbf{V},\mathbf{A})$ and V$_\text{ortho} \equiv \psi_a \psi_b f_2(\mathbf{V})$ with the same features of eqs.~\ref{eq_flip} and~\ref{eq_cmcm} (see SI~\cite{supp}, section V).
Additionally, several $A(B,B')_2$X$_4$ spinels may adopt cation-ordered structures of $P4_{1,3}22$ symmetry ~\cite{Talanov2014-ik} (e.g. LiZnNbO$_4$~\cite{Marin1994, Liu2019, PhysRevLett.105.075501}, $B/B'$ ratio = 1:1), a process suitable to be influenced by V$_\text{flip}$ and V$_\text{ortho}$ due to the X$_3$ character of the occupational order parameter.

Realizing a reversible and controllable enantiomeric excess is therefore interlaced with the direct activation of polar and axial modes. 
The absence of a true conjugate field for the chiral order indicates that any handedness-flipping interaction can activate achiral orthorhombic competing degrees of freedom as well.
While it is possible for a symmetry mode to be polar and chiral at the same time~\cite{PhysRevB.109.024113}, so that the chirality and associated gyrotropic properties can be controlled by an electric field~\cite{iwasaki1971,Iwasaki1972} as in the case of Pb$_{5}$Ge$_{3}$O$_{11}$~\cite{fava2023ferroelectricity}, 
this requires the axial mode to be invariant~\cite{fava2023ferroelectricity} which is not the case for enantiomorphic crystals such as KNO.
In practice realizing an enantiomeric excess (or "chirality poling") could be achieved via chiral phonon-light coupling and, possibly, by working close to the transition temperature via adiabatic cooling. 
By symmetry, the $\mathbf{T}$ tensor coupling with the Z soft eigenmode may take the form of the so-called "Lipkin zilch" ~\cite{PhysRevLett.104.163901}, which defines the chirality density - $\rho_{\chi} = \frac{\epsilon_{0}}{2}\calE\cdot(\mathbf{\nabla}\times\calE) + \frac{1}{2\mu_{0}}\calB\cdot(\mathbf{\nabla}\times\calB)$ - of a circularly polarised electromagnetic (EM) field. 
While no detailed study of an enantiomeric excess realized in the presence of a chiral phonon instability has been reported, the control over the phonon handedness in quartz has recently been obtained with the aid of circularly polarized X-rays\cite{Ueda2023}.

%%%%%%%%%%%%%%%%%%%%%%%%%%%%%%%%%%%%%%
\textit{Conclusions}.
We have observed \textit{ab initio} how the chirality in crystals can emerge as a displacive spontaneous order triggered by a degenerate soft mode. By using the K$_3$NiO$_2$ compound as a model system, we have shown how its symmetry is lowered from $P4_2/mnm$ to the enantiomorphic $P4_12_12$, $P4_32_12$ pair by a soft Z$_4$ mode which can be controlled by epitaxial strain and hydrostatic pressure.
Finally, we have numerically explored the possible realization of a stable enantiomeric excess mediated by a coupling between chiral, polar, and axial orders.
Our model applies as long as order parameters with opposite handedness belong to the same representation, a condition encountered in a large variety of systems.
Importantly, we have shown that the $E_g$ modes, when activated via polar-polar-Raman coupling~\cite{PhysRevLett.118.054101}, cannot efficiently lift the right/left degeneracy and favor an orthorhombic distortion ($mmm$ point group) instead.
Hence, realizing a permanent handedness conversion could more likely be achieved by using an axial conjugate field on top of an electric field.
Additional analysis would be appealing to understand how the gyrotropic physical properties of crystals are affected by this perspective on chirality.

\section*{Acknowledgements}
Computational resources have been provided by the Consortium des \'Equipements de Calcul Intensif (C\'ECI), funded by the Fonds de la Recherche Scientifique (F.R.S.-FNRS) under Grant No. 2.5020.11.
MF \& EB acknowledges FNRS for support and the PDR project CHRYSALID No.40003544. Work at West Virginia University was supported by the U.S. Department of Energy (DOE), Office of Science, Basic Energy Sciences (BES) under Award DE‐SC0021375. This work used Bridges2 and Expanse at the Pittsburgh Supercomputer (Bridges2) and the San Diego Supercomputer Center (Expanse) through allocation DMR140031 from the Advanced Cyberinfrastructure Coordination Ecosystem: Services \& Support (ACCESS) program, which National Science Foundation supports grants 2138259, 2138286, 2138307, 2137603, and 2138296.

\bibliography{bibliography}

\newpage

\section{Supplementary Material}

\subsection{Details of the calculations}

Electronic computations and structural relaxation on the K$_{3}$NiO$_{2}$ compound have been performed with the help of the VASP (v5.4) code~\cite{Kresse1999} adopting a cutoff of 550 eV and a 6x6x6 kpoint grid. The PBE functional~\cite{Perdew1996} and a tolerance of 10$^{-6}$ eV over the self-consistent convergence has been used (unless otherwise specified in the text). The Hubbard-U correction (with the scheme devised by Dudarev \textit{et al.}~\cite{PhysRevB.57.1505}) has been used and with the reference value U = 4.2 eV found through the method of Cococcioni et al.~\cite{PhysRevB.71.035105}.
The initial magnetic moments on the Nickel sites is +1 $\mu_\text{B}$, which is found stable upon optimization.
Finally, structural analysis has been performed with the help of ISODISTORT~\cite{isodistort}
and AMPLIMODES~\cite{amplimode}.
The phonon spectrum has been calculated with the help of a 2x2x2 supercell (96 ions) and the Phonopy software~\cite{phonopy1,phonopy2}.

\subsection{High-symmetry soft phonon modes and Born effective charges, chiral DOS}

\begin{table}[htbp!]
\begin{center}
\begin{tabular}{ccccc}
\hline
\hline
 Mode  & freq.(THz) & $\Delta$E (meV) \rule{0pt}{8pt}\\
\hline
Z & 1.35i & 8.9  \rule{0pt}{8pt}\\
$\Gamma$ & 0.80i & 3.0 \\
M & 0.80i & 4.2 \\
R & 1.15i & 3.6 \\
X & 0.45i & 3.7 \\
\hline
\end{tabular}
\caption{Soft modes of KNO ($P4_2/mnm$). $\Delta$E$_\text{mode}$ $\equiv$ E($P4_2/mnm$) - E(mode) is given formula unit. We use U = 4.2 eV throughout the calculations. We also report $\Delta$E(U = 3.2 eV) = 7.7 meV f.u. and $\Delta$E(U = 5.2 eV) = 10.4 meV f.u..}
\label{tab:soft_modes}
\end{center}
\end{table}

\begin{table}[htbp!]
\begin{center}
\begin{tabular}{ccccc}
\hline
\hline
 Atom  & site sym. &  &  &  \rule{0pt}{8pt}\\
\hline
Ni$_1$ & $2a$ & 
$\begin{pmatrix}
 0.25  &  1.66 &  0.00 \\
 1.66  &  0.25 &  0.00 \\
 0.00  &  0.00 &  -1.10 
 \end{pmatrix}
 $
  &  
 
 &
  \\
K$_1$ & $4d$ & 
$\begin{pmatrix}
 1.06 &   -0.04 &  0.00 \\
 0.04 &   1.06 &  0.00 \\
 0.00 &   0.00 &  1.10
\end{pmatrix}
$
& 

&  \\
K$_2$ & $2b$ &  
$
\begin{pmatrix}
 1.34 & -0.22 &  0.00 \\
-0.22 & 1.34 &  0.00 \\
 0.00 & 0.00 &  1.19
\end{pmatrix}
$
& 

&  \\
O$_1$ & $4f$ &  
$
\begin{pmatrix}
 -1.86 &  -0.84 & 0.00 \\
-0.84 &  -1.86 & 0.00 \\
 0.00 &  0.00 & -1.15 
\end{pmatrix}
$
& 

&  
 \\

\hline
\end{tabular}
\caption{Calculated Born effective charges of nonequivalent Ni, K and O ions in the P4$_{2}$/mnm high symmetry phase.}
\label{tab:BEC}
\end{center}
\end{table}

\begin{figure}
     \includegraphics[width=0.5\textwidth]{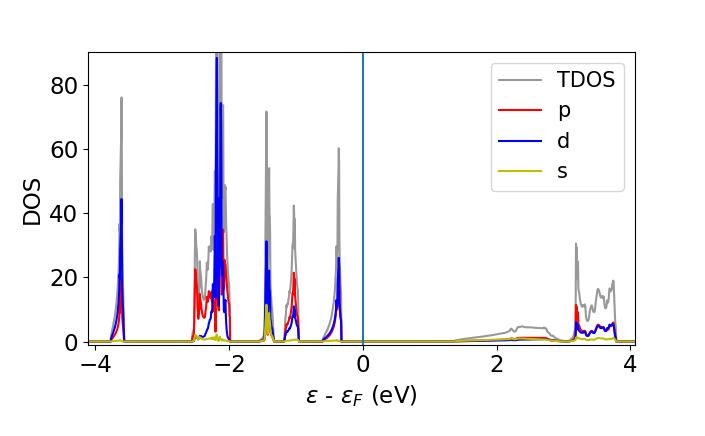}
     \caption{Partial density of states of K$_3$NiO$_2$ in the $P4_{1}$2$_{1}$2 (or $P4_{3}$2$_{1}$2) phase near the Fermi level. The $s$ and $p$ weights are largely associated with oxygen while the $d$ corresponds mainly to Ni (VB) or K (CB).}
     \label{fig:pdos_low}
     \end{figure}

In table~\ref{tab:soft_modes} we show the soft modes of K$_3$NiO$_2$. The (fully relaxed) energy gain (formula unit) with respect to the high symmetry $P4_2/mnm$ phase are reported.
In the case of the $\Gamma$ ($E_u$) instability, $\Delta$E($\Gamma$) = 2.3 meV  (3.8 meV) in presence of a 3 \% tensile (compressive) strain.
The Born effective charges (BEC) of KNO are instead reported in tab.~\ref{tab:BEC}. Their non-anomalous features suggest that the polar $E_u$ state stabilized under tensile strain (see main text) is originated by short-range forces.
Finally, the density of states in the chiral phase is shown in fig.~\ref{fig:pdos_low}. Much like in the achiral reference, the VBM is dominated by O-$2p$ / Ni-$3d$ hybridization while the CBM contains a significant potassium-$d$ orbitals contribution.

\subsection{About the Z$_4$ and A$_{1u}$ symmetry modes}

The Z$_4$ main order parameter, upon structural relaxation, can be described by: (i) an in-plane helical displacement (0.4 \AA/ion) of the K(2) ions (which belong to the Ni-O-K(2)-O linear chains and localize at the $2b$ site in the $P4_2/mnm$ phase~\cite{djurivs2012k3nio2}, as already shown in main figure 1) along with a smaller motion of (ii) the tetrahedral K(1) ions (0.17 \AA/ion, from the $4d$ site in the $P4_2/mnm$ phase) and (iii) the oxygen atoms (0.14 \AA/ion).

\begin{figure}
\centering
     \includegraphics[width=0.46\textwidth]{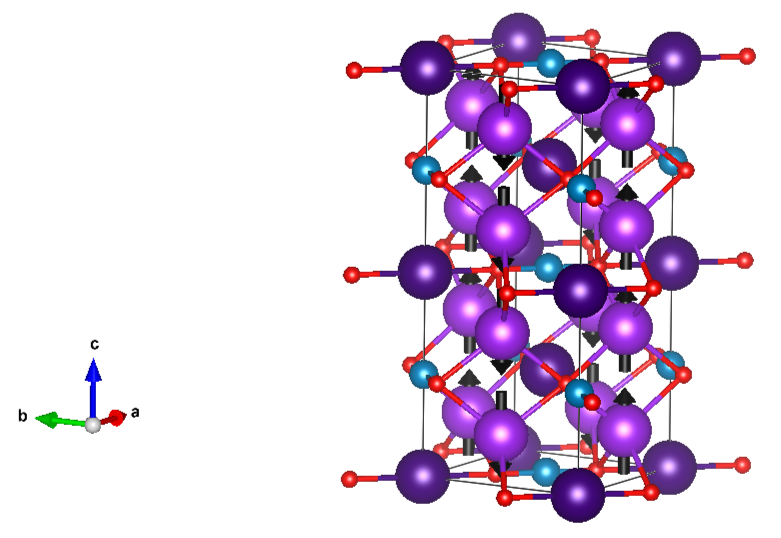}
     \centering
     \caption{$A_{1u}$ mode (the black arrows describe atomic displacements) superimposed to a (1,1,2) supercell of the high symmetry phase. Oxygen and nickel ions are the red and cyan spheres. We have used light and dark purple spheres to represent the K(1) and K(2) Wyckoff positions respectively. While the Z$_4$ mode mostly involves an in-plane spiral displacement of the K(2) ions, the pseudoscalar SAM only involves a small out-of-plane motion (0.006 \AA/ion) of the K(1) atoms at the centre of KO$_4$ tetrahedral units.}
    \label{fig:A1u}
     \end{figure}

The $A_{1u}$ secondary deformation has the space group $P4_22_12$ (no. 94, which is achiral) as the associated isotropy subgroup. A depiction of this secondary distortion with respect to the $P4_2/mnm$ phase can be observed in fig.~\ref{fig:A1u}. In particular, while the Z$_4$ SAM is mostly in-plane, the pseudoscalar mode only involves a tiny displacement (0.006 \AA/ion) along the $c$ axis of the K(1)-ions from the centres of the K(1)O$_4$ tetrahedra, with half the K(1) ions moving upwards and half moving downwards.

\subsection{Chiral-axial-polar coupling}\label{symm}

\subsubsection{Application of an electric field to the Na$_{3}$AuO$_{2}$"toy model" system: procedure}

We have used the ABINIT software v9.6.2~\cite{gonze_2020} and the PAW approximation (JTH v1.1) via PseudoDojo project~\cite{pseudodojo} (v0.4). The GGA functional (PBE) has also been adopted along with a 6x6x6 kpoint grid and a cutoff of 22 and 44 Ha on the external and internal grids respectively.

We have taken the KNO unit cells and replaced K with Na and Ni with Au while preserving the symmetries (and the asymmetric units) of K$_{3}$NiO$_{2}$ to give a Na$_{3}$AuO$_{2}$ "toy model" isostructural to KNO. The high symmetry $P4_{2}/mnm$ has then been relaxed (both ions and cell parameters) and the cell parameters used as a reference for the low symmetry phase alike.
To be noted that all the low-symmetry (chiral and $Cmcm$) structures are taken in P1 setting and with the same lattice parameters to accommodate the polar and axial distortions.

Every ion-only structural optimizations in the presence of an electric field (4.63x10$^{4}$ kV/m along the [1,1,1] direction) have been performed on the $P4_{2}/mnm$ or on the $Pnnm$ ($P4_{2}/mnm$ in presence of an additional 1 \% $xx$-strain) phases. The corresponding axial $E_g$ and polar ($E_u$, $A_{2u}$) modes have been extracted and applied unaltered to the energy surface (as a function of the Z$_4$ amplitude). 

Finally, we remark that while the $\phi_{a}\phi_{b}\phi_{A_{2u}}\phi_{A_{2g}}$ term is also symmetry allowed and takes a role analogous to that of $V_\text{flip}$, we have  not considered it since the $A_{2g}$ representation is silent.

\subsubsection{Energy surface in the presence of strain}

\begin{figure}
     \includegraphics[width=0.5\textwidth]{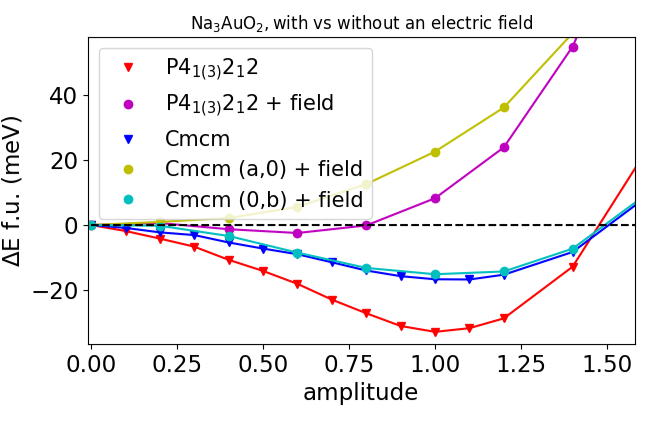}
     \caption{Energy landscape of Na$_{3}$AuO$_{2}$ as a function of the Z mode amplitude, with/without an electric field [1,1,1] x 4.63x$10^{4}$ kV/m. The relevant distortions are labelled by their respective isotropy subgroups.}
     \label{fig:w_field_no_strain}
     \end{figure}

\begin{figure}
     \includegraphics[width=0.45\textwidth]{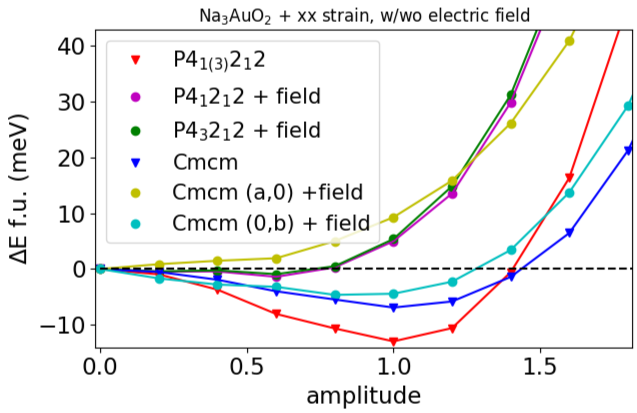}
     \caption{Energy landscape of Na$_{3}$AuO$_{2}$ as a function of the Z mode amplitude, with/without an electric field [1,1,1] x 4.63x$10^{4}$ kV/m and with a 1 \% strain along $x$. The relevant distortions are labelled by their respective isotropy subgroups.}
     \label{fig:w_field_w_strain}
     \end{figure}

The effects of an electric field generated polar + Raman modes  
acting on the $P4_2/mnm \rightarrow P4_{1(3)}2_12$ energy surface are shown in fig.~\ref{fig:w_field_no_strain}. One can see that the $Cmcm$ (0,b) $Z_4$ distortion is favoured at the present field strength magnitude while no right/left degeneracy splitting can occur because $V_{\text{Raman-polar}} \sim \phi_{A_{2u}}(\phi^{a}_{R}\phi^{b}_{IR} - \phi^{b}_{R}\phi^{a}_{IR})$ induces the vanishing of $V_\text{flip}$.
This issue can in principle be rectified
by applying a strain $\mathbf{\epsilon}$ along one of the in-plane axes (e.g. $\epsilon_{xx}$ along the $a$ axis), so that $4/mmm \rightarrow mmm$.  
Upon this further condition and with considering that the polar responses of KNO and NAO are paraelectric in nature, namely $\phi_{IR}\sim\calE_{xy}$ and $\phi_{A_{2u}}\sim\calE_{z}$, we can write:

\begin{equation}
    V_\text{flip} = \alpha(\epsilon_{xx})\calE_{xy}^2\calE_{z}\phi_a\phi_b,
\end{equation}

with the coefficient $\alpha\rightarrow{0}$ in the $\epsilon_{xx}\rightarrow{0}$ limit. Nevertheless, this effect (measured by the energy difference between $P4_{1}$2$_1$2 and $P4_{3}$2$_1$2 mirror-symmetric configurations) is proven to be small compared to the effect of the $V_{\text{ortho}}\sim(\phi_a^2-\phi_b^2)\phi^{a}_{IR}\phi^{b}_{IR}$ invariant as reported in fig.~\ref{fig:w_field_w_strain}, where the DFT-computed $\Delta$E of Na$_{3}$AuO$_{2}$ is shown as a function of the soft Z mode and with $\epsilon_{xx}$ = 1\%.

\subsection{Enantioselective mechanism in hexagonal perovskites and spinels}

Chiral enantiomorphic order parameters can emerge in several classes of materials from a single degenerate instability of the energy landscape. Examples of these materials include hexagonal perovskites and spinels. We will now discuss how the enantioselective mechanism proposed in this work could manifest in these systems.

\subsubsection{Hexagonal perovskites}

$ABX_{3}$ hexagonal halide perovskites (A = Rb, Cs, B = V, Cr, Mn, Fe, Co, Ni, Cu, and X = F, Cl, Br, I) are known to realize spin chains~\cite{Yamamoto2021} and are also of interest for their optical, electronic and chemical properties~\cite{Yamamoto2021}. As mentioned in the main text, CsCuCl$_{3}$ ~\cite{hirotsu1977} undergoes a $P6_3/mmc\rightarrow P6_122, P6_522$ transition with a Jahn-Teller-triggered instability at the $\Delta = (0,0,1/3)$ point of the Brillouin zone. In this case, the polar (A$_{2u}$, E$_u$) and axial (A$_{2g}$, E$_{g}$) modes at $\Gamma$ are irreducible representations of the $6/mmm$ point group. At the same time, the $\Delta$ soft mode belongs to the $\Delta_6$ irrep. The corresponding order parameter takes the form $(a,b,0,0)$ (I take the last two components to be zero for simplicity of treatment, plus $a,b$ $\rightarrow \Delta_a, \Delta_b$). Here, $\Delta_b$ = 0, $\Delta_a\neq$ 0 identifies the $P6_522$ phase while $\Delta_a$ = 0, $\Delta_b\neq$ 0 its enantiomorph $P6_122$. The direction ($\Delta_a$,$\pm\Delta_a$) identifies a $Cmcm$ distortion and a generic ($\Delta_a$,$\Delta_b$) direction again has $C222_1$ symmetry group. The energy invariants that are relevant to reproduce in this class of materials the enantioselective mechanism studied throughout this work are as follows:

\begin{itemize}
    \item $A_{2u}(E^{a}_{u}E^{b}_{g}-E^{b}_{u}E^{a}_{g})$ which is the Raman-polar-polar interaction. The axial E$_g$ modes can be activated via the simultaneous application of an in-plane and an out-plane electric field;
    \item $(\Delta_{a}^2 - \Delta_{b}^2)A_{2g}A_{2u}$ and $(\Delta_{a}^2 - \Delta_{b}^2)(E^{a}_{u}E^{a}_{g} + E^{b}_{u}E^{b}_{g})$ which play the role of $V_\text{flip}$ since they break the $P6_122$-$P6_522$ degeneracy and favor one enantiomorph over the other;
    \item $\Delta_a\Delta_b({E^{a}_{u}}^2 - {E^{b}_{u}}^{2})$ which plays  the role of $V_\text{ortho}$ by favouring a $Cmcm$ distortion over each enantiomorphic pair. Moreover and like V$_\text{ortho}$ in KNO, this term breaks the degeneracy between ($\Delta_a,\Delta_a$) and ($\Delta_a$,-$\Delta_a$) orthorhombic deformations.
\end{itemize}

As in the case of KNO, the Raman-polar interaction when used to switch the Raman E$_g$ modes on results into V$_\text{flip}$ = 0 and V$_\text{ortho} \neq$ 0 and is thus unsuited to generate a permanent enantiomeric excess. As in K$_3$NiO$_2$ in the main text, a direct control of the axial degrees of freedom (attainable through the application of an axial field) is thus necessary to overcome the $P6_122\rightarrow Cmcm \rightarrow P6_522$ (or the $P6_122\rightarrow P6_3/mmc \rightarrow P6_522$) barrier.

\subsubsection{Spinels}

The 
displacive mechanism for an achiral – chiral phase transition 
in spinels is the orbital-driven displacive transition observed for MgTi$_2$O$_4$, as also mentioned in the main text. On cooling below 260 K, MgTi$_2$O$_4$ undergoes a first-order phase transition to an enantiomorphic phase of $P4_12_12$ (or $P4_32_12$) symmetry as pairs of 3d$^1$ Ti$^{3+}$ cations dimerize, giving a metal-insulator transition.~\cite{Isobe2002,Schmidt2004}

This $Fd\Bar{3}m\rightarrow P4_32_12, P4_12_12$ phase transition in MgTi$_2$O$_4$ is triggered by a soft X$_4$ (0,1,0) mode. 
The polar and axial symmetry modes belong to the (triply degenerate) T$_{1u}$ and T$_{1g}$ irreducible representations of the parent cubic $m\Bar{3}m$ point group. The X$_4$ mode is also degenerate 
and we take the direction $(a,b;0,0,0,0)$ in representation space. As before we relabel $a,b$ $\rightarrow$ X$_a$, X$_b$. Here X$_a$ = 0, X$_b$ $\neq$ 0 (X$_b$ = 0, X$_a$ $\neq$ 0) identifies the distortion with $P4_12_12$ ($P4_32_12$) symmetry. X$_a$ = $\pm$ X$_b$ describes a $Pnma$ phase instead and a generic (X$_a$,X$_b$) state has $P2_12_12_1$ symmetry group.
Due to the cubic nature of the parent group, we can identify the following invariants that reproduce the enantioselective mechanism explored in the present letter:

\begin{itemize}
    \item $T^{a}_{1u}T^{b}_{1u}T^{a}_{1g} + T^{a}_{1u}T^{c}_{1u}T^{c}_{1g} + T^{b}_{1u}T^{c}_{1u}T^{b}_{1g}$ which is the Raman-polar-polar interaction (the superscripts label the three cartesian components);
    \item $({X^2_a} - {X^2_b})(T^{a}_{1u}T^{b}_{1g} -T^{c}_{1u}T^{a}_{1g})$ which acts like $V_\text{flip}$ since it removes the $P4_12_12$-$P4_32_12$ degeneracy and favors one enantiomorphic state over the other;
    \item $X_a X_b (T^a_{1u} T^c_{1u})$ favors a $Pnma$ distortion over the chiral ground state and also breaks the (X$_a$,X$_a$) - (X$_a$,-X$_a$) degeneracy.
    Therefore this term is the MgTi$_{2}$O$_{4}$-equivalent of $V_\text{ortho}$.
\end{itemize}

As for the previous example materials, we can see that the biquadratic activation of the axial modes results in V$_\text{flip}$ = 0 and V$_\text{ortho}\neq$ 0. Hence, again, 
an axial conjugate field is required to generate a stable enantiomeric excess.

\end{document}